\documentclass[11pt]{article}
\usepackage[margin=1in]{geometry}
\usepackage{amsmath}
\usepackage[english]{babel}
\usepackage{graphicx}
\usepackage{natbib}
\usepackage{array}
\usepackage{rotating}
\usepackage{color}
\usepackage{setspace}
\usepackage{multirow}
\usepackage{float}

\begin{document}
\begin{center}{\Large {\bf Optimal two-treatment crossover designs\\ for binary response models}}\end{center}
\vspace{.5em}

\begin{center}{{\sc By SIULI MUKHOPADHYAY$^1$, SATYA PRAKASH SINGH}\\
{\it Department of Mathematics, Indian Institute of Technology Bombay,}\\ {\it Mumbai 400 076, India}\\siuli@math.iitb.ac.in, spsingh@math.iitb.ac.in}\end{center}
\vspace{.25em}

\begin{center}{{\sc and ALOKE DEY}\\
{\it Theoretical Statistics \& Mathematics Unit,}\\ {\it Indian Statistical Institute, New Delhi 110 016, India}\\aloke.dey@gmail.com}\end{center}
\date{}
\footnote{Corresponding author}
\doublespacing

\begin{center}
{\sc Summary}\end{center}

Optimal two-treatment, $p$ period crossover designs for binary responses are determined. The optimal designs are obtained by minimizing the variance of the treatment contrast estimator over all possible allocations of $n$ subjects to $2^p$ possible treatment sequences. An appropriate logistic regression model is postulated and the within subject covariances are modeled through a working correlation matrix. The marginal mean of the binary responses are fitted using generalized estimating equations. The efficiencies of some crossover designs for $p=2,3,4$ periods are calculated. The effect of misspecified working correlation matrix on design efficiency is also studied.  
\vspace{.5em}

\noindent {\it Some key words}: Binary response; Generalized estimating equations; Logistic regression; Efficiency.
\vspace{.5em}

\begin{center}{1. {\sc Introduction}}\end{center}

Crossover trials, wherein every experimental subject is exposed
to a sequence of treatments over different periods of time, have been applied in a variety of
areas; see for example Jones \& Kenward (2014) and Senn (2003) for real life examples. The problem of determining optimal designs for crossover trials has been studied quite extensively in recent years and we refer to Bose \& Dey (2009) for a review of results on optimal crossover designs.  Most of the available results on optimal crossover designs relate to situations where the response variable is continuous. However, there are situations in practice where the response in a crossover trial is binary in nature. For example, consider a trial reported by Senn (2003, page 127) wherein it was desired to study the effect of two drugs on 24 children aged 7 to 13 suffering from exercise-induced asthma. The two treatments were, a single dose of 12$\mu g$ formoterol solution  aerosol, treatment A and a single dose of 200$\mu g$ of salbutamol solution aerosol, treatment B.  Each child was given both the treatments either in the order, AB or BA. The response variable was binary in nature taking value 1 if the drug was effective and $0$ otherwise. An equal number of children were allocated to each treatment sequence, AB or BA. Is this the best design to be used in this situation?

In clinical or pharmaceutical research, the outcome of interest is often binary in nature. While methods for analysing binary data arising from crossover trials are available in Jones \& Kenward (2014) and Senn (2003), the question of designing such studies in an optimal manner does not seem to have been addressed. Waterhouse et al. (2006) considered crossover designs for binary response, where the treatments were taken to be continuous in nature and no period effects were considered in the model. 

In this article, optimal crossover designs are studied when the response variable is binary. We consider crossover trials with two treatments and $p$ periods. The proposed designs minimize the variance of the estimator of treatment contrast of direct effects over all possible allocation of the $n$ subjects to the treatment sequences. In the logistic regression model considered, both direct effect as well as the carryover effect of each treatment are considered, wherein we assume that the carryover effect of a treatment lasts only to the next succeeding period. While analysing data from binary crossover trials, often it is assumed that all observations are mutually uncorrelated; however this is not a very realistic assumption. We therefore assume that the $p$ observations from each subject are mutually correlated while the observations from different subjects are uncorrelated. The correlation between observations within subjects are modeled using a ``working correlation structure". Since the main interest is in estimating the treatment effects, we treat the subject effects as a nuisance parameter and use the generalized estimating equations of Liang \& Zeger (1986) to estimate the marginal means. Though estimating equations were used earlier by Jones \& Kenward (2014) to analyse repeated measures data, their models did not include the carryover effect. 

The variance of the treatment effect estimator depends on the model parameters. To address the issue of parameter dependence, local optimal designs are found for given values of the model parameters. For $p=2,3,4$, we study  the effect of two working correlation structures, equi-correlated and autoregressive (AR) on the designs chosen.  We also look at the effect of misspecification of the covariance  on the design efficiency.

In \S 2, we define the crossover logistic model for a binary response and discuss the estimation of the crossover model using generalized estimating equations. In \S 3, results on optimal two-treatment designs for 2, 3 and 4 periods are given. 

\begin{center}{2. {\sc The model and estimation}}\end{center}

Consider a crossover trial involving $t$ treatments, $n$ subjects and $p$ periods. Suppose the response obtained from the $j$th subject is  $Y_j=(Y_{1j},\ldots,Y_{pj})'$, where a prime denotes transposition.  Instead of specifying a joint distribution of the repeated measurements we use a working generalized linear model (GLM) to describe the marginal distribution of $Y_{ij}$ as in Liang \& Zeger (1986),
\begin{equation*}
f(y_{ij})=exp[\{y_{ij}{\phi_{ij}}-b(\phi_{ij})+c(y_{ij})\}\psi],
\end{equation*}
For a binary random variable $Y_{ij}$, $\displaystyle \phi_{ij}=\log\frac{\mu_{ij}}{1-\mu_{ij}}$, $b(\phi_{ij})=\log[1+\exp\{\phi_{ij}\}]$, $c(y_{ij})=0$, and the scale parameter $\psi$ is 1 (Robinson \&  Khuri, 2003).
The mean of $Y_{ij}$ is $\mu_{ij}$ and variance $\mu_{ij}(1-\mu_{ij})$.

In a crossover setup, we model the marginal mean $\mu_{ij}$ as
\begin{equation}
logit (\mu_{ij})=\eta_{ij}=\mu+\beta_i+\tau_{d(i,j)}+\rho_{d(i-1,j)};\,i=1\,\ldots,p,\,j=1,\ldots,n,
\end{equation}
where
$\mu$ is the overall mean, $\beta_i$ represents the effect of the $i$th period, $\tau_s$ is the direct effect due to treatment $s$ and $\rho_s$ is the carryover effect due to treatment $s$, , $s=1,\ldots ,t$. Throughout, $1_u$ is a $u\times 1$ vector of all ones, $I_u$ is the identity matrix of order $u$ and $0_{ab}$ is an $a\times b$ null matrix. Also, we write $\beta=(\beta_1,\ldots,\beta_p)'$,
$\tau=(\tau_1,\ldots,\tau_t)'$  and $\rho=(\rho_1,\ldots,\rho_t)'$.  

 In matrix form the linear predictor corresponding to the $j$th subject, $\eta_j=(\eta_{1j},\ldots,\eta_{pj})'$, can be written as
\begin{equation}
\eta_j=X_j\theta,
\end{equation}
where $\theta=(\mu,\beta,\tau,\rho)'$. The design matrix is $X_j =[1_{p} \;P_j \;T_j\;F_j]$, where
$P_j=I_p$; $T=(T'_1,\ldots,T'_n)'$, where $T_j$ is a $p\times t$ matrix with its $(i,s)$th entry equal to 1 if subject $j$ receives the direct effect of the treatment $s$ in the $i$th period and zero otherwise; $F=(F'_1,\ldots,F'_n)'$, where $F_j$ is a $p\times t$ matrix with its $(i,s)$th entry equal to 1 if subject $j$ receives the carryover effect of the treatment $s$ in the $i$th period and zero otherwise.

The estimating equations of Liang and Zeger (1986) and Zeger et al. (1988) are used to estimate the regression coefficients and to obtain their variances. It is assumed that measurements from the same subject in the $p$ periods are correlated while observations from different subjects are  uncorrelated.

The dependencies between repeated observations from a subject are modeled using a "working correlation" matrix $R(\alpha)$ where $\alpha$ is a vector of length $s$. If $R(\alpha)$ is the true correlation matrix of $Y_j$, then
\begin{equation}
Cov[Y_j]=A_j^{1/2}R(\alpha)A_j^{1/2},
\end{equation}
where $A_j=diag(\mu_{1j}(1-\mu_{1j}),\ldots,\mu_{pj}(1-\mu_{pj}))$. Let $V_j=A_j^{1/2}R(\alpha)A_j^{1/2}$.

For a repeated-measures model, Zeger et al. (1988, equation (3.1)) define the generalized estimating equations (GEE) to be
\begin{equation*}
\sum_{j=1}^{n}\frac{\partial\mu'_j}{\partial\theta}V_{j}^{-1}(Y_j-\mu_j)=0,
\end{equation*}
where $\mu_j=(\mu_{1j},\ldots,\mu_{pj})'$. The asymptotic variance for the GEE estimator $\hat{\theta}$ (see Zeger et al. 1988, equation (3.2)) is
\begin{equation}
Var(\hat{\theta})=
\left[\sum_{j=1}^{n}\frac{\partial\mu'_j}{\partial\theta} V_{j}^{-1}\frac{\partial\mu_j}{\partial\theta}\right]^{-1},
\end{equation}
if $Cov(Y_j)=V_j$. However, if the true correlation structure varies from the ``working correlation" structure, then $Var(\hat{\theta})$ is given by the sandwich formula (Zeger et al. 1988, equation (3.2))
\begin{equation}
Var(\hat{\theta})=
\left[\sum_{j=1}^{n}\frac{\partial\mu'_j}{\partial\theta} V_{j}^{-1}\frac{\partial\mu_j}{\partial\theta}\right]^{-1}\left[\sum_{j=1}^{n}\frac{\partial\mu'_j}{\partial\theta} V_{j}^{-1}Cov(Y_j)V_{j}^{-1}\frac{\partial\mu_j}{\partial\theta}\right]\left[\sum_{j=1}^{n}\frac{\partial\mu'_j}{\partial\theta} V_{j}^{-1}\frac{\partial\mu_j}{\partial\theta}\right]^{-1}.
\end{equation}

For the crossover model (1), the $i$th element of $\frac{\partial\mu_j}{\partial\theta}$  is $\frac{\partial\mu_{ij}}{\partial\theta}=x'_{ij} \mu_{ij}(1-\mu_{ij})$, where $x'_{ij}$ is the $i$th row of $X_j$ for $i=1,\ldots,p$.

For finding optimal crossover designs for the logistic model we use the approximate theory as in Laska \& Meisner (1985) and Kushner (1997, 1998). For a review of results on optimal crossover designs using the approximate theory, we refer to Bose \& Dey (2009, Chapter 4). Fixing the number of subjects to $n$ and  periods to $p$, we determine the proportion of subjects assigned to a particular treatment sequence.
Each treatment sequence is of length $p$ and a typical sequence can be written as
$\omega=(t_1,\ldots,t_p)', \,t_i\in \{1,\ldots,t\}$. Let ${\Omega}$ be the set of all such sequences. We denote by $n_\omega$ the number of subjects assigned to sequence $\omega$. Then, $n=\sum_{\omega\in\Omega}n_\omega,n_\omega\geq 0$. A design $\zeta$ in approximate theory is specified by the set $\{p_\omega,\omega\in\Omega\}$ where $p_\omega=n_\omega/n$, is the proportion of subjects assigned to treatment sequence $\omega$.

Note that the matrices $T_j$ and $F_j$ depend only on the treatement sequence $\omega$ to which the $j$th subject is assigned,  so $T_j=T_\omega,\,F_j=F_\omega$; see Lemma 4.2.1 in Bose \& Dey (2009). This implies, $X_j=X_\omega$. Since $np_\omega$ subjects are assigned to treatment $\omega$, the variance of $\hat{\theta}$ is
\begin{equation}
Var_\zeta(\hat{\theta})=\left[\sum_{\omega\in\Omega}np_\omega\frac{\partial\mu'_\omega}{\partial\theta} V_{\omega}^{-1}\frac{\partial\mu_\omega}{\partial\theta}\right]^{-1}\left[\sum_{\omega\in\Omega}np_\omega\frac{\partial\mu'_\omega}{\partial\theta} V_{\omega}^{-1}Cov(Y_\omega)V_{\omega}^{-1}\frac{\partial\mu_\omega}{\partial\theta}\right]\left[\sum_{\omega\in\Omega}np_\omega\frac{\partial\mu'_\omega}{\partial\theta} V_{\omega}^{-1}\frac{\partial\mu_\omega}{\partial\theta}\right]^{-1}.
\end{equation}
As before, if the true correlation of $Y_j$ is equal to $R(\alpha)$ then \begin{equation}
Var_\zeta(\hat{\theta})=\left[\sum_{\omega\in\Omega}np_\omega\frac{\partial\mu'_\omega}{\partial\theta} V_{\omega}^{-1}\frac{\partial\mu_\omega}{\partial\theta}\right]^{-1}.
\end{equation}

In crossover trials the main interest usually lies in estimating the direct treatment effect contrasts. Thus, instead of working with the full variance-covariance matrix of $\hat{\theta}$ we concentrate on $Var(\hat{\tau})$  where,
\begin{equation}
Var_\zeta(\hat{\tau})=EVar_\zeta(\hat{\theta})E',
\end{equation}
where $E$ is a $t\times m$ matrix given by $[0_{t1},0_{tp},I_t,0_{tt}]$ and $m$ is the total number of parameters in $\theta$.  

An optimal design is one which minimizes the variance of $\hat{\tau}$ over the set of all possible allocations of the $n$ subjects to the $2^p$ treatment sequences.


\begin{center}{3. \sc {Two treatment crossover trials: results and discussion}}\end{center}

The optimality and efficiency of two-treatment crossover designs for continuous responses have been studied by various authors, including Kershner \& Federer (1981), Laska \& Meisner (1985), Matthews (1987) and Carriere \& Huang (2000). With two treatments of interest, the problem simplifies to minimizing the variance of the treatment contrast $\tau_1-\tau_2$ to obtain optimal crossover designs. Reparametrizing model (1) as in Laska \& Meisner (1985), using $\tau=(\tau_1-\tau_2)/2$ and $\rho=(\rho_1-\rho_2)/2$ we get
\begin{equation}
logit(\mu_{ij})=\mu+\beta_i+\tau \Phi_{d(i,j)}+\rho\Phi_{d(i-1),j)},
\end{equation}
where $\Phi_A=1$,$\Phi_B=-1$ and $\Phi_{d(0,j)}=0$.

For illustration we go back to the example in \S 1, where there are two treatments, A and B applied in two periods to each child. The design used involved the treatment sequences $AB$ and $BA$, with equal allocation to each treatment sequence. Thus, the matrix $X_\omega$ depends on the treatment sequence $\omega\in\Omega=\{AB,BA\}$. If the treatment sequence, for example is $ \omega=\{AB\}$, then
\begin{equation*}
X_\omega=\left( \begin{array}{cccccccc}
1 & 1 & 0&1&0\\
1 & 0 & 1&-1&1\end{array} \right)
\end{equation*}
In the following, we look at the performance of the design, $\{AB,BA\}$. Senn (2003) fitted a logistic model with no carryover effect to the data set and computed confidence intervals for the various components of $\theta$. Using these intervals we investigate if the above two-period design is the best choice in the given situation. We also look at general situations for determining optimal designs when $p=2,3$ or 4 for the two treatment case. 
\vspace{.25em}

\noindent 3.1 {\it  Designs compared}. The designs that we consider are the same as those discussed by Laska \& Meisner (1985) and Carriere \& Huang (2000) and are listed below for $p=2,3$ and 4:

\noindent (i) $p=2$:

\noindent Design 1: $AB$ and $BA$; Design 2: $AB$, $AA$, $BA$ and $BB$,
with equal number of subjects assigned to each sequence. 

\noindent (ii) $p=3$:  

\noindent  $d_1$: $ABB$ and $BAA$, with optimal allocation to each treatment sequence;

\noindent $d_2$: $ABB$ and $BAA$;

\noindent $d_3$: $ABB$, $AAA$, $BAA$ and $BBB$;

\noindent $d_4$: $ABB$, $AAB$, $BAA$ and $BBA$;

\noindent $d_5$: $ABB$, $ABA$, $BAA$ and $BAB$;

\noindent $d_6$: $AAA$ and $BBB$.

\noindent In designs $d_2-d_6$, each treatment sequence is allocated equally.

\noindent (iii) $p=4$:

\noindent I: $AABB$, $BBAA$, $ABBA$, $BAAB$, with optimal allocation to each treatment sequence;

\noindent  II: $AABB$, $BBAA$, $ABBA$ and $BAAB$;

\noindent III: $AABB$, $BBAA$;

\noindent IV: $ABBA$,  $BAAB$.

\noindent In designs II-IV, each treatment sequence is allocated equally.

For evaluating and comparing the above designs we define an efficiency measure as
\begin{equation}
Eff(\zeta)=\left(\frac{Var_{\zeta^*}(\hat{\tau})}{Var_{\zeta}(\hat{\tau})}\right)^{1/m},
\end{equation}
where $\zeta^*$ is the optimal crossover design obtained. 
\vspace{.25em} 

\noindent 3.2. {\it Working correlation structures}. We consider the uncorrelated, compound symmetric or, equi-correlated and the AR(1) structures for the correlation matrix $R(\alpha)$.
Under the equi-correlated covariance structure, 
$R_j=(1-\alpha)I_p+\alpha J_p$.

Under the AR(1) assumption,
\begin{equation*}
R_j=\alpha^{|i-i'|},\,i\neq i'
\end{equation*}

\vspace{.25em}

\noindent 3.3 {\it Results}. We begin with the design for the trial reported in \S 1. A model without the carryover effect is fitted to the binary data. The parameter estimates and confidence intervals for the parameters are: $\mu\in[1.1573,5.0893]$; $\tau\in[-5.1738,-1.4029]$; $\beta\in[-2.8390,0.4932]$, as listed in Senn (2003).  A  parameter space for $\hat{\theta}$ is set up by  taking Cartesian product of the confidence intervals. Under uncorrelated error structure, the efficiency of both designs 1 and 2 are the same.  For the equi-correlated error structure, we take $\alpha=0.2, 0.4$ for finding the best design. The performance of the design $\{AB,BA\}$ is studied through the distribution of the efficiency values in the parameter space and compared with the design $\{AB,BA,AA,BB\}$. During comparisons $10,000$ values are randomly selected from the parameter space and the efficiencies computed. It is found that design 1 is substantially superior to design 2 in terms of median and minimum efficiencies. Furthermore, the efficiency of design 1 is less affected by an increase in the value of $\alpha$.

We next determine optimal crossover designs for $p=2,3,4$ under the correlation structures stated earlier. 
For logistic models, the variance of $\hat{\tau}$ depends on the parameter values. To address the issue of parameter dependence we use various parameter spaces, $\mathcal{B}_1-\mathcal{B}_6$. For an explanation of the different parameter spaces considered, see Table \ref{partable}. For instance, suppose $p=2$ and the parameter space is $\mathcal{B}_1$. Then,  $\mu\in[-0.5,0.5]$, $\beta_1\in[-1,1],$ $\beta_2\in[-1,1]$, $\tau\in[-1.5,1.5]$, $\rho\in[-1,1]$ and $\alpha=0.2$.  For each value of $p$, the designs  listed in subsection 3.1 are compared by choosing $10,000$ values from each of the six parameter spaces $\mathcal{B}_1-\mathcal{B}_6$. The performance of these designs are examined by noting their minimum and median efficiencies. The length of the parameter intervals considered in $\mathcal{B}_1-\mathcal{B}_3$ are more when compared to those from $\mathcal{B}_4-\mathcal{B}_6$. This enables us to study the effect of increasing  parameter uncertainty on the chosen designs.

\noindent 3.3.1. {\it Estimation of direct treatment effects in the presence of carryover effects}.
When the carryover effect is included in the model and $p=2$, we only consider design 2 for estimation of direct treatment effects, because, as noted by Bose \& Dey (2015), design 1 does not permit the estimation of contrasts among direct effects. First, let us consider the uncorrelated error structure. All designs for $p=2,3$ and $4$ are compared over parameter spaces $\mathcal{B}_1$ and $\mathcal{B}_4$.  It is seen that design 2 has high median efficiency ($99\%$) for $p=2$ (see Table \ref{effp2t2}).

For $p=3$ the distribution of efficiencies of the designs are presented in Table \ref{efftablep3t2}. From this table we note that design $d_4$ performs the best followed closely by designs $d_1$ and $d_2$. Design $d_6$ is the worst. The performance of design $d_4$ is not much affected by increasing lengths of parameter intervals. Under uncorrelated errors for $p=4$, design I performs the best in terms of median and minimum efficiencies (see Table \ref{efftablep4t2}). On the average design I allocates equal proportion of subjects to each treatment sequence. 

We next consider a compound symmetric correlation structure for the errors. All designs were compared over six parameter spaces, $\mathcal{B}_1-\mathcal{B}_6$. For a model including carryover effects and $p=2$, design 2 has $99\%$ median efficiency (see Table \ref{effp2t2}). Thus, as in the case of continuous responses (Laska \& Meisner, 1985), design 2 performs well even under a logistic model. Design 2 is also not much affected by changes in the correlation parameter and in the lengths of the parameter intervals.

 The distribution of efficiencies of the designs for $p=3$ are presented in Table \ref{efftablep3t2}.  From the table we see that with respect to median efficiency, the design $d_1$ with optimal allocation of subjects to each sequence is the best, very closely followed by designs $d_2$ and $d_4$. The median efficiencies of these three designs are also not affected by increasing $\alpha$. The minimum efficiency values of design $d_4$ are the highest. The effect of increasing $\alpha$ is more on the minimum efficiencies values of designs $d_1$ and $d_2$ as compared to $d_4$. Thus, on the basis of both median and minimum efficiencies, the performance of design $d_4$ is the best among the six designs compared. Design $d_6$ has the worst efficiency values and its performance is affected by increasing values of $\alpha$. Note that under the continuous response case, $d_2$ with equal allocation to sequences ABB and BAA is known to be the universally
optimal design within the class of three-period designs for equi-correlation (Laska \& Meisner, 1985).

For $p=4$, median efficiencies of all designs I-IV are at least $98\%$, and do not change with increasing $\alpha$ (see Table \ref{efftablep4t2}). However, when designs are compared with respect to minimum efficiencies, design I is the best closely followed by design II, unchanged with changes in the correlation. Both designs III and IV record lower minimum efficiency values, which decrease with increasing $\alpha$.  On the average design I allocates $n/4$ subjects to each sequence.  For the normal linear model case, Laska \& Meisner (1985) showed that design II is optimal.

Finally we consider the AR(1) correlation structure. For $p=3$, designs $d_1,d_2$ and $d_4$ record at least $99\%$ median efficiency values (see Table \ref{efftablep3t2}). However, the minimum efficiencies of designs $d_1$ and $d_2$ are lower than those of design $d_4$ and are also affected by increasing correlation. Even for the AR(1) structure, we see that the best design is $d_4$ on the basis of both minimum and median efficiencies and it is also least affected by increasing $\alpha$. Design $d_6$, with equal allocation to sequences AAA and BBB is the worst among the designs compared.   Laska \& Meisner (1985) showed that for $p=3$ in the continuous response case, design $d_4$ is optimal. However for all $\alpha$ more individuals were allocated to sequence AAB and its dual.

Under the AR(1) structure, for $p=4$, design I is the best design with highest median and minimum efficiencies. Designs III and IV do not perform well, their minimum efficiencies are lower and decrease with increasing $\alpha$ (see Table \ref{efftablep4t2}).The average allocations of design I is studied, the results show that for low values of $\alpha$, the sequence $ABBA$ and its dual each gets $35\%$ allocation while, sequence $AABB$ and its dual get $15\%$ allocation. As $\alpha$ increases to $0.4$, $85\%$ individuals are allocated to  $ABBA$ and its dual, while for $\alpha=0.6$, it is $92\%$. Thus, as $\alpha$ increases almost no subjects are  allocated to  sequence $AABB$ and its dual. Our result matches with those of Laska \& Meisner (1985) and Matthews (1987) when the response is continuous.

\noindent 3.3.2. {\it Estimation of carryover effects}.
Although direct treatment effects are generally of primary interest in crossover trials, we present some results on optimal designs for estimating carryover effects as well for the uncorrelated, equicorrelated and AR(1) structures. For $p=3,4$,  the parameter spaces considered are $\mathcal{B}_1-\mathcal{B}_3$. 

In the uncorrelated error case for $p=3$, we note that $d_4$ performs the best with $99.51\%$ median and $96.23\%$ minimum efficiencies. The median efficiencies of $d_1$ and $d_2$ are also at least $99\%$; however their minimum efficiencies ($90\%$) are much lower than those under $d_4$ ($96\%$). 

 For the equicorrelated and AR(1) covariance structures,  $d_1$ and $d_2$ have $99\%$ median efficiencies and at least $90\%$ minimum efficiency values. Also the performance of designs $d_1$ and $d_2$ do  not change with increasing correlation. However, unlike in the estimation of direct treatment effect the performance of $d_4$ is not good and is affected by changes in $\alpha$. Median efficiency of $d_4$ dcreases to $94\%$ and minimum efficiency to $90\%$  as $\alpha$ increases to $0.6$. Design $d_6$ is the worst among the designs compared as before. On the average  $d_1$ allocates equal ($n/2$) subjects to sequence $ABB$ and its dual. For the normal case under AR(1) covariance structure, Matthews  (1987) showed that $d_2$ has $100\%$ efficiency for $\alpha=0.2,0.4,0.6$.

When $p=4$, under all three covariance structures, design I performs the best with respect to minimum and median efficiencies. Under uncorrelated and compound symmetric covariances, the average allocation of design I is $n/4$ to each sequence. However, under AR(1), as $\alpha$ increases more subjects get allocated to sequence $AABB$ and its dual than sequence $ABBA$ and its dual. In the normal response case under AR(1) covariance structure, Matthews (1987)  showed that design with sequences AABB, ABBA and their duals  has at least $85\%$ efficiency for positive $\alpha$.
\vspace{.25em}

\noindent 3.4. {\it Effect of covariance misspecification}

So far we have assumed that the true correlation structure of the responses is equal to the working correlation structure. However, this may not be true in most cases. To see the effect of varying structures on the efficiency of the designs, we carry out a simple study  We set $p=3$ and use parameter spaces $\mathcal{B}_5$ and $\mathcal{B}_6$. The working correlation structure is taken to be compound symmetric and the true correlation AR(1). The designs studied are $d_1,d_2$ and $d_4$. 
The distribution of efficiencies of the three designs when working and true structures are assumed to be equal are already given in Table 3.

 For finding the efficiency of the designs when there is misspecification we use (6) in the variance formula and report the results in Table \ref{misefftablep3t2}. We see that the performance of all designs are affected; however designs $d_1$ and $d_2$ suffer much more than $d_4$. In case of $\alpha=0.4$, the median efficiencies of designs $d_1$ and $d_2$ reduce to $97\%$,  and to $94\%$ when $\alpha$ is $0.6$. The efficiency of design $d_4$ is at least $97\%$ for $\alpha=0.4, 0.6$.  The minimum efficiencies of designs $d_1$ and $d_2$ are affected much more than design $d_4$ and also more when we consider a higher value of $\alpha$. Thus, design $d_4$ appears to be the best design to use when we have carryover effects in the model as it  also guards against misspecification in covariance.
\vspace{.25em}

\noindent 3.5. {\it Concluding Remarks}.
Crossover designs for binary responses are compared for $p=2,3,4$. Since these designs depend on the parameter values, intervals of the parameters are considered and local optimal designs are found in each case. The main results on the estimation of direct effects are summarized below.

For $p=3$ design $d_4$ is seen to be the best design for uncorrelated,  equicorrelated and AR(1) covariance structures. For $p=4$ design I is seen to be the most efficient design for uncorrelated, equicorrelated and AR(1) covariance structures. In the equicorrelated case, on average design I allocates equal number of subjects to the treatment sequences. However, in the AR(1) case, the average optimal allocation depends on the correlation parameter $\alpha$. The results found in the logistic regression case for $p=2,3,4$ are very similar to available results in the continuous case. 

\vspace{.5em}

\begin{center}
{\sc Acknowledgement}\end{center}

The work of A. Dey was supported by the National Academy of Sciences, India under the Senior Scientist scheme of the Academy. The support is gratefully acknowledged. \\
The work of S. Mukhopadhyay was supported by Department of Science and Technology, India[Grant Number: SR/FTP/MS-
13/2009]. The support is gratefully acknowledged.
\vspace{.5em}

\begin{center}{\sc References}\end{center}
\begin{description}
\item {\sc Bose, M.} \& {\sc Dey, A.} (2009). {\it Optimal Crossover Designs.} Singapore: World Scientific.
\item {\sc Bose, M.} \& {\sc Dey, A.} (2015). Crossover designs. In {\it Handbook of Design and Analysis of Experiments} (A. M. Dean, M. Morris, J. Stufken, D. Bingham, Eds.), Chapman \& Hall/CRC Press, pp. 159--95.
\item {\sc Carriere, K. C.} \& {\sc Huang, R.} (2000). Crossover designs for two-treatment clinical trials. {\it J. Statist. Plan. Infer.} {\bf 87}, 125--34.
\item {\sc Jones, B.} \& {\sc Kenward, M. G.} (2014). {\it Design and Analysis of Crossover Trials}, 3rd ed. London: CRC Press.
\item {\sc Kershner, R. P.} \& {\sc Federer, W. T.} (1981). Two-treatment crossover designs for estimating a variety of effects. {\it J. Amer. Statist. Assoc.} {\bf 76}, 612--19.
\item {\sc Kushner, H. B.} (1997). Optimal repeated measurements designs: the linear optimality equations. {\it Ann. Statist.} {\bf 25}, 2328--44.
\item {\sc Kushner, H. B.} (1998). Optimal and efficient repeated measurements designs for uncorrelated observations. {\it J. Amer. Statist. Assoc.} {\bf 93}, 1176--87.
\item {\sc Laska, E. M.} \& {\sc Meisner, M.} (1985). A variational approach to optimal two-treatment crossover designs: application to carryover-effect models. {\it J. Amer. Statist. Assoc.} {\bf 80}, 704--10.
\item {\sc Liang, K. Y.} \& {\sc Zeger, S. L.} (1986). Longitudinal data analysis using generalized linear models. {\it Biometrika} {\bf 73}, 13--22.
\item {\sc Matthews, J. N. S.} (1987). Optimal crossover designs for the comparison of two treatments in the presence of carryover effects and autocorrelated errors. {\it Biometrika} {\bf 74}, 311--20.
\item {\sc Robinson, K. S.} \& {\sc Khuri, A. I.} (2003). Quantile dispersion
                  graphs for evaluating and comparing designs for logistic
                  regression models. {\it Comput. Statist. Data Anal.} {\bf 43}, 47--62.
\item {\sc Senn, S.} (2003). {\it Cross-over Trials in Clinical Research}, 2nd ed. Chichester, England: Wiley.
\item {\sc Waterhouse, T. H.}, {\sc Eccleston, J. A.} \& {\sc Dufull, S. B.} (2006). Optimal crossover designs for logistic regression models in pharmacodynamics. {\it J. Biopharm. Statist.} {\bf 16}, 881--94.
\item {\sc Zeger, S. L.}, {\sc Liang, K. Y.} \& {\sc Albert, P. S.} (1988). Models for longitudinal data: a generalized estimating equation approach. {\it Biometrics} {\bf 44}, 1049--60.
\end{description}
\newpage
\begin{table}[p]
\caption{Parameter spaces}
\begin{center}
\begin{tabular}{ccccccccc}
\hline\\
Parameters & \mbox{}& $\mu$ & $\beta_i$ (for each $i$ = 1, 2, 3, 4)  & $\tau$ & $\rho$ & $\alpha$\\
\hline\
\multirow{6}{*}{Range} &$\mathcal{B}_1$& [-0.5, 0.5] & [-1.0, 1.0] &[-1.5, 1.5]& [-1.0, 1.0]&0.2\\
&$\mathcal{B}_2$&[-0.5, 0.5] & [-1.0, 1.0]&[-1.5, 1.5]& [-1.0, 1.0]&0.4\\
&$\mathcal{B}_3$&[-0.5, 0.5] & [-1.0, 1.0]&[-1.5, 1.5]& [-1.0, 1.0]&0.6\\
&$\mathcal{B}_4$& [-0.5, 0.5] & [-1.0, 1.0]&[-0.2, 1.5]& [-0.2, 1.0]&0.2\\
&$\mathcal{B}_5$&  [-0.5, 0.5] & [-1.0, 1.0]&[-0.2, 1.5]& [-0.2, 1.0]&0.4\\
&$\mathcal{B}_6$&[-0.5, 0.5] & [-1.0, 1.0]&[-0.2, 1.5]& [-0.2, 1.0]&0.6\\
\hline
\end{tabular}\label{partable}
\end{center}
\end{table}

\begin{table}[p]
\caption{Efficiencies (minimum, median) of  design  ($D_2$)   compared to the optimal design $D^{*}$ for 2-periods 2-treatments  with carry-over effect model. Correlation Structures (Corr. st.) are compound symmetric (CS) and uncorrelated (Ind.).}
\begin{center}
\begin{tabular}{ccccc}
\hline\\
\multirow{1}{*}{Par. Space} &Corr. St.& $D_2$\\
\hline\
\multirow{2}{*}{$\mathcal{B}_1$}&CS&(0.9639,    0.9977)\\
\mbox{}&Ind.&(0.9599,    0.9974)\\

$\mathcal{B}_2$&CS&(0.9539,    0.9975)\\
$\mathcal{B}_3$&CS&(0.9595,    0.9972)\\
\multirow{2}{*}{$\mathcal{B}_4$}&CS&(0.9738,    0.9983)\\
\mbox{}&Ind.&(0.9664,    0.9977)\\
$\mathcal{B}_5$&CS&(0.9749,    0.9981)\\
$\mathcal{B}_6$&CS&(0.9734,    0.9980)\\
\hline
\end{tabular}\label{effp2t2}
\end{center}
\end{table}

\begin{sidewaystable}[p]
\caption{Estimation of direct treatment effect. Efficiencies (minimum, median) of  designs $d_1-d_6$ as  compared to the optimal design $D^{*}$ for 3-periods 2-treatments model. Correlation Structures (Corr. st.) are compound symmetric (CS), AR(1) and uncorrelated (Ind.).}
\begin{center}
\begin{tabular}{ccccccccc}
\hline\\
 \mbox{}&Par. Space&  Corr. St.   & $d_1$ & $d_2$ & $d_3$& $d_4$& $d_5$& $d_6$\\
\hline\\\\
&\multirow{3}{*}{$\mathcal{B}_1$}&CS&(0.9364,    0.9972)&(0.9355,    0.9964)&(0.9135,    0.9487)&(0.9744,    0.9972)&(0.9382,    0.9668)&(0.7514,    0.8139)\\
&\mbox{}&AR(1)&(0.9381,    0.9945)&(0.9374,    0.9938)&(0.9244,    0.9636)&(0.9704,    0.9960)&(0.9409,    0.9714)&(0.7629,    0.8318)\\
&\mbox{}&Ind.&(0.9434,    0.9960)&(0.9423,    0.9952)&(0.9304,    0.9686)&(0.9742,    0.9979)&(0.9390,    0.9682)&(0.7720,    0.8378)\\\\

&\multirow{2}{*}{$\mathcal{B}_2$}&CS&(0.9293,    0.9974)&(0.9291,    0.9966)&(0.8964,    0.9308)&(0.9707,    0.9959)&(0.9337,    0.9658)&(0.7255,    0.7796)\\
&\mbox{}&AR(1)&(0.9309,    0.9917)&(0.9309,    0.9910)&(0.9054,    0.9503)&(0.9657,    0.9939)&(0.9350,    0.9708)&(0.7438,    0.8141)\\\\
&\multirow{2}{*}{$\mathcal{B}_3$}&CS&(0.8955,    0.9971)&(0.8955,    0.9963)&(0.8830,    0.9156)&(0.9659,    0.9943)&(0.9247,    0.9650)&(0.6803,    0.7315)\\
&\mbox{}&AR(1)&(0.9027,    0.9860)&(0.9027,    0.9853)&(0.8813,    0.9290)&(0.9579,    0.9912)&(0.9195,    0.9662)&(0.6978,    0.7749)\\\\
&\multirow{2}{*}{$\mathcal{B}_4$}&CS&(0.9501,    0.9982)&(0.9497,    0.9976)&(0.9294,    0.9511)&(0.9813,    0.9975)&(0.9396,    0.9645)&(0.7565,    0.8138)\\
&\mbox{}&AR(1)&(0.9467,    0.9961)&(0.9461,    0.9954)&(0.9417,    0.9666)&(0.9821,    0.9970)&(0.9415,    0.9690)&(0.7680,    0.8328)\\
&\mbox{}&Ind.&(0.9501,    0.9973)&(0.9489,    0.9966)&(0.9446,    0.9725)&(0.9853,    0.9986)&(0.9403,    0.9651)&(0.7699,    0.8375)\\\\

&\multirow{2}{*}{$\mathcal{B}_5$}&CS&(0.9408,    0.9984)&(0.9407,    0.9977)&(0.9132,    0.9326)&(0.9776,    0.9958)&(0.9392,    0.9636)&(0.7271,    0.7795)\\
&\mbox{}&AR(1)&(0.9310,    0.9937)&(0.9308,    0.9931)&(0.9233,    0.9533)&(0.9773,    0.9953)&(0.9369,    0.9689)&(0.7478,    0.8160)\\\\
&\multirow{2}{*}{$\mathcal{B}_6$}&CS&(0.9201,    0.9983)&(0.9200,    0.9975)&(0.8910,    0.9166)&(0.9735,    0.9941)&(0.9290,    0.9631)&(0.6838,    0.7314)\\
&\mbox{}&AR(1)&(0.9036,    0.9881)&(0.9034,    0.9875)&(0.8894,    0.9317)&(0.9672,    0.9929)&(0.9158,    0.9647)&(0.7016,    0.7777)\\\\
\hline
\end{tabular}\label{efftablep3t2}
\end{center}
\end{sidewaystable}

\begin{table}[p]
\caption{Estimation of direct treatment effect. Efficiencies (minimum, median) of  designs $I-IV$ compared to the optimal design $D^{*}$ for 4-periods 2-treatments model. Correlation Structures (Corr. st.) are compound symmetric (CS), AR(1) and uncorrelated (Ind.).}
\begin{center}
\begin{tabular}{ccccccccc}
\hline\\\\
 \mbox{}&Par. Space&  Corr. St.& $D_{I}$ & $D_{II}$ & $D_{III}$& $D_{IV}$\\
\hline\\\\
&\multirow{3}{*}{$\mathcal{B}_1$}&CS&(0.9803,    0.9996)&(0.9787,    0.9982)&(0.9151,    0.9848)&(0.9228,    0.9845)\\
&\mbox{}&AR(1)&(0.9772,    0.9994)&(0.9746,    0.9962)&(0.9046,    0.9728)&(0.9315,    0.9927)\\
&\mbox{}&Ind.&(0.9767,    0.9985)&(0.9759,    0.9970)&(0.9180,    0.9840)&(0.9245,    0.9838)\\\\

&\multirow{2}{*}{$\mathcal{B}_2$}&CS&(0.9805,    0.9996)&(0.9792,    0.9983)&(0.9000,    0.9845)&(0.9164,    0.9842)\\
&\mbox{}&AR(1)&(0.9754,    0.9999)&(0.9678,    0.9906)&(0.8816,    0.9571)&(0.9204,    0.9972)\\\\

&\multirow{2}{*}{$\mathcal{B}_3$}&CS&(0.9733,    0.9996)&(0.9707,    0.9982)&(0.8894,    0.9837)&(0.8860,    0.9837)\\
&\mbox{}&AR(1)&(0.9720,    1.0000)&(0.9592,    0.9829)&(0.8591,    0.9404)&(0.8975,    0.9986)\\\\

&\multirow{3}{*}{$\mathcal{B}_4$}&CS&(0.9878,    0.9997)&(0.9847,    0.9986)&(0.9504,    0.9876)&(0.9186,    0.9811)\\
&\mbox{}&AR(1)&(0.9870,    0.9996)&(0.9837,    0.9973)&(0.9369,    0.9761)&(0.9275,    0.9911)\\
&\mbox{}&Ind.&(0.9875,    0.9989)&(0.9855,    0.9977)&(0.9520,    0.9872)&(0.9228,    0.9812)\\\\

&\multirow{2}{*}{$\mathcal{B}_5$}&CS&(0.9917,    0.9997)&(0.9840,    0.9987)&(0.9368,    0.9872)&(0.8999,    0.9810)\\
&\mbox{}&AR(1)&(0.9807,    1.0000)&(0.9770,    0.9926)&(0.9163,    0.9607)&(0.9151,    0.9969)\\\\

&\multirow{2}{*}{$\mathcal{B}_6$}&CS&(0.9899,    0.9997)&(0.9834,    0.9988)&(0.9306,    0.9868)&(0.8859,    0.9809)\\
&\mbox{}&AR(1)&(0.9720,    1.0000)&(0.9644,    0.9852)&(0.8835,    0.9448)&(0.9030,    0.9988)\\\\
\hline
\end{tabular}\label{efftablep4t2}
\end{center}

\end{table}

\begin{table}[p]
\caption{Efficiencies  (minimum, median) of  designs $d_1, d_2$ and $d_4$ under misspecification of correlation structure.}
\begin{center}
\begin{tabular}{ccccccccc}
\hline\\
Par. Space& $d_1$ & $d_2$& $d_4$\\
\hline\
$\mathcal{B}_5$&(0.9124,    0.9716)&(0.9120,    0.9710)
&(0.9651,    0.9821)\\
 $\mathcal{B}_6$&(0.8713, 0.9456)&(0.8709, 0.9448)
&(0.9456, 0.9670)\\
\hline
\end{tabular}\label{misefftablep3t2}
\end{center}
\end{table}
\end{document}